\documentclass[english,prb,twocolumn,showpacs,
superscriptaddress,floatfix,longbibliography]{revtex4-1}

\usepackage[T1]{fontenc}
\usepackage[utf8]{inputenc}
\usepackage{amsmath,amssymb,amstext}
\usepackage{xcolor}
\usepackage{graphicx}
\usepackage{esint}
\usepackage{babel}
\usepackage{verbatim}
\usepackage{color}
\usepackage{subfigure}
\usepackage{braket}
\usepackage[pdfborder={0 0 1}]{hyperref}
\usepackage{ulem}
\newcommand{\udarr}{{\uparrow(\downarrow)}}

\begin{document}

\title{Gap inversion in quasi-one-dimensional Andreev crystals}

\author{Mikel Rouco} \email{mikel.rouco@ehu.eus} \affiliation{ Centro de
  F\'isica de Materiales (CFM-MPC), Centro Mixto CSIC-UPV/EHU, Manuel de
  Lardizabal 5, E-20018 San Sebasti\'an, Spain }

\author{F. Sebastian Bergeret} \email{fs.bergeret@csic.es} \affiliation{ Centro
  de F\'isica de Materiales (CFM-MPC), Centro Mixto CSIC-UPV/EHU, Manuel de
  Lardizabal 5, E-20018 San Sebasti\'an, Spain } \affiliation{ Donostia
  International Physics Center (DIPC), Manuel de Lardizabal 4, E-20018 San
  Sebastian, Spain }

\author{Ilya V. Tokatly} \email{ilya.tokatly@ehu.es} \affiliation{ Nano-Bio
  Spectroscopy group, Departamento F\'isica de Materiales, Universidad del
  Pa\'is Vasco, Av. Tolosa 72, E-20018 San Sebasti\'an, Spain } \affiliation{
  IKERBASQUE, Basque Foundation for Science, E48011 Bilbao, Spain }
\affiliation{ Donostia International Physics Center (DIPC), Manuel de Lardizabal
  4, E-20018 San Sebastian, Spain }

\date{\today}

\begin{abstract}
  We study a periodic arrangement of magnetic regions in a quasi-one-dimensional
  superconducting wire. Due to the local exchange field, each region supports
  Andreev bound states that hybridize forming Bloch bands in the subgap spectrum
  of what we call the Andreev crystal (AC). As an illustration, ACs with
  ferromagnetic and antiferromagnetic alignment of the magnetic regions are
  considered. We relate the spectral asymmetry index of a spin-resolved
  Hamiltonian to the spin polarization and identify it as the observable that
  quantifies the closing and reopening of the excitation gap. In particular,
  antiferromagnetic ACs exhibit a sequence of gapped phases separated by gapless
  Dirac phase boundaries. Heterojunctions between antiferromagnetic ACs in
  neighboring phases support spin-polarized bound states at the interface. In a
  close analogy to the charge fractionalization in Dirac systems with a mass
  inversion, we find a fractionalization of the interface spin.
\end{abstract}

\maketitle

\section{Introduction}

\begin{figure}
    \centering
    \includegraphics[width=.9\linewidth]{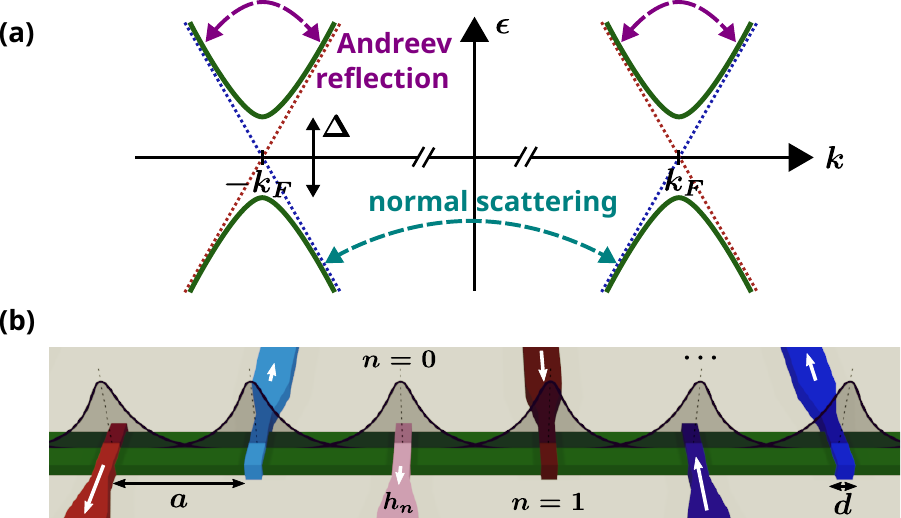}
    \caption{(a) Schematic drawing of the specturm of a quasi-1D s-wave superconductor. The electron- (dashed blue line) and hole-branches (dashed red line) couple forming valleys (solid green line) and opening a gap equal to $\Delta$ at the Fermi surface. As sketched by the dashed arrows, Andreev reflections couple electron- and hole-branches within each valley, whereas normal back-scattering events couple quasiparticles at different valleys. (b) Sketch of a possible experimental realization of an Andreev
    crystal: A superconducting wire (in green) is contact to ferromagnetic
    fingers. The latter induce a local exchange field in the superconductor in
    the direction of the white arrows, strong enough to locally break the superconducting phase. The black curve above the structure
    represents the localized Andreev states bounded to each magnetic region
    which hybridize forming the Andreev bands.}
    \label{fig:intro-sketches}
\end{figure}

Non-magnetic impurities\footnote{regions of much smaller size than the superconducting coherence length, $\xi_0$.}  in a superconductor do not modify substantially its spectrum. In contrast,  a magnetic defect may lead to bound states localized around this region\cite{yu-1965-bound, shiba-1968-classical, rusinov-1968-superconductivity,
  andreev-1966-electron, sakurai-1970-comments, yazdani-1997-probing,
  balatsky-2006-impurity, franke-2011-competition, rouco-2019-spectral,
  heinrich-2018-single, farinacci-2018-tuning}. 
The features of such bound states depend on the size of magnetic impurity and the strength of exchange interaction\cite{rouco-2019-spectral}.  
In a quasi-one-dimensional (quasi-1D) ballistic superconducting wire\footnote{a wire which lateral dimensions are much smaller than the superconducting coherence length.} with a magnetic region one can distinguish two different limiting cases. 
In one case the magnetic exchange
coupling is strong and concentrated at a point-like impurity, resuting into the appearence of two non-degenerate states within the superconducting gap with opposite energies with respect to the Fermi energy. These are the so-called
Yu-Shiba-Rusinov (YSR) states\cite{yu-1965-bound, shiba-1968-classical, rusinov-1968-superconductivity}. 
In the other limiting case the magnetic region has a finite size and the exchange energy is small compared to the Fermi energy. Electrons can go through the magnetic region without being back-scattered. Instead, they can only  be reflected as holes via the  Andreev reflection \cite{andreev-1966-electron}.  Such reflection events couple the electron- and hole-branches within each valley at the $\pm k_F$ points (see sketch in Fig.~\ref{fig:intro-sketches}a) and induces two pairs of degenerate  bound
states in the superconducting gap with opposite energies, $\pm\epsilon_0$, known as Andreev bound states. This degeneracy can be lifted if the two Fermi valleys  are coupled via back-scattering.

Whereas YSR-states are generated by magnetic impurities of atomic size, Andreev bound states can be found in ballistic mesoscopic magnetic  regions\cite{konschelle-2016-ballistic,rouco-2019-spectral}. This latter case is well described from a semiclassical perspective\cite{konschelle-2016-semiclassical} in which electrons, crossing the magnetic region, accumulate a
spin-dependent phase, $e^{i\sigma\Phi}$. Here $\Phi = \int \frac{dx}{\hbar v_F}
h(x)$ is the phase accumulated for a collinear exchange field $h(x)$, the Fermi velocity is $v_F$, and
$\sigma=\pm$ encodes the difference between spin up/down electrons. Holes accumulate the same phase, but
with opposite sign, $e^{-i\sigma\Phi}$.

When several impurities form a  periodic chain, it is natural to expect that  single-impurity bound states hybridize forming bands.  Such bands  have been widely studied in chains of magnetic atoms in the YSR-limit\cite{nadj-perge-2013-proposal, pientka-2013-topological, heimes-2014-majorana, poyhonen-2014-majorana, weststrom-2015-topological, pientka-2015-topological, brydon-2015-topological, schecter-2016-self}, mainly motivated by the  appearance of  topological phases that may  host Majorana zero-modes on the endings of the wire.  However, little attention has been paid to chains of semiclassical magnetic impurities that can be, for example, realized in a mesoscopic superconducting wire connected to a periodic array of ferromagnetic electrodes, as sketeched in Fig.~\ref{fig:intro-sketches}b.

In this article we study these semiclassical chains, which we denote  as \textit{Andreev crystals} (ACs). We focus on AC  with a collinear magnetization and analyze its spectral properties and possible quantum phases that emerges by changing the parameters of the chain. 
Phases are gapped and separated by gapless regions. 
We identify the total spin of the system as the observable which reveals these different phases. As discussed in section~\ref{sec:spin-asym-index},   from a very general perspective,  the spin is determined by the asymmetry index of the spin-resolved Hamiltonian,
{\it i.e.} the difference between the number of states below and above the Fermi
energy. In gapped systems this index, and hence the spin, can only change by closing the gap.  In section~\ref{sec:andreev-crystals} we focus on two types of ACs, ferromagnetic and
antiferromagnetic chains, and determine the corresponding spectra.  By changing
the magnetic phase, $\Phi$, we find gapped and gapless phases.  In particular,
antiferromagnetic ACs exhibit  a sequence of gapped phases separated, at half-integer values of $\Phi/\pi$, by gapless phase boundaries with Dirac
points. 
In section~\ref{sec:AC-junctions} we show that a hybrid system with two such semi-infinite antiferromagnetic ACs
may exhibit spin-polarized bound states at the interface, which are similar to
the states found in Dirac systems with a spatial mass inversion
\cite{jackiw-1976-solitons, su-1980-solitonb, su-1979-solitons,
  volkov-1985-two}.

\section{Spin polarization in systems with collinear exchange fields}
\label{sec:spin-asym-index}

We consider a quasi-1D s-wave superconductor\footnote{We focus on the one-dimensional
  case to be consistent with the rest of the paper, but one can derive the
  expression for the total spin of a system of any dimensions following the same
  lines.} in the presence of a collinear exchange field, $\hat h(x) =
\hat\sigma_z h(x)$, such that the spin along the $z$ direction is a conserved
quantity (here, $\hat\sigma_z$ stands for the third Pauli matrix). The
Bogoliubov-de Gennes (BdG) Hamiltonian describing the system is block diagonal
in spin with
\begin{equation}
  \label{eq:spec-asy-single-spin-BdG-hamiltonian}
  \hat H_\sigma(x) = \hat\tau_3 \xi + \hat\tau_1\Delta(x) - \sigma h(x).
\end{equation}
Here, $\hat\tau_{i=1,2,3}$ are the Pauli matrices spanning the Nambu
(electron-hole) space, $\hat\xi$ stands for the quasiparticle energy operator,
$\Delta(x)$ describes the superconducting order parameter, and $\sigma$ is the
spin label \footnote{The spin label for spin up (down) is substituted by $\sigma
  = \uparrow (\downarrow)$ when it appears as a subscript, whereas it takes
  values of $\sigma = +(-)$ when it is part of an equation.}.  From the
corresponding imaginary-frequency Green's functions (GFs), $\hat G_\udarr
(\epsilon) = [i\epsilon - \hat H_\udarr ]^{-1}$, one can compute the total spin
polarization of the system at zero temperature:
\begin{equation}
  \label{eq:spec-asy-spin-polarization-T0}
  S = \frac{\hbar}{4} \lim_{\tau \rightarrow 0} \text{Tr}
  \int \frac{d\epsilon}{2\pi} \Big[\hat G_\uparrow(\epsilon) - \hat G_\downarrow(\epsilon)
  \Big] e^{i\epsilon\tau} ,
\end{equation}
where the trace runs over the coordinate$\,\otimes\,$Nambu space. Since the spin-up and -down components of the Hamiltonian are related by the transformation $\hat
H_\downarrow = -\hat\tau_2 \hat H_\uparrow \hat\tau_2$, then the GFs also fulfill that $\hat
G_\downarrow(\epsilon) = -\hat\tau_2 \hat G_\uparrow(-\epsilon) \hat\tau_2$.
Substituting this relation into Eq.~\eqref{eq:spec-asy-spin-polarization-T0} and
using the cyclic property of the trace we obtain:
\begin{align}
  \frac{2S}{\hbar}
  & = \frac{1}{2} \lim_{\tau \rightarrow 0} \text{Tr} \int \frac{d\epsilon}{2\pi}
  \bigg[\frac{1}{i\epsilon - \hat H_\uparrow} + \frac{1}{-i\epsilon - \hat H_\uparrow}\bigg] e^{i\epsilon\tau}
  \nonumber \\
    \label{eq:spec-asy-spin-polarization-as-spectral-asym}
  & = - \frac{1}{2} \lim_{\tau \rightarrow 0} \sum_n \text{sgn}(E_{n\uparrow}) e^{-|E_{n\uparrow}|\tau},
\end{align}
where $E_{n\uparrow}$ stands for the energy of the $n$-the eigenstate of the
spin-up Hamiltonian. The expression in the last line corresponds to the difference between the
number of states below and above the Fermi energy for a given spin projection,
and it is known as the spectral asymmetry index, widely used in topology
\cite{atiyah-1976-spectralIII, atiyah-1975-spectralII,atiyah-1975-spectralI},
quantum field theory and condensed matter
physics\cite{niemi-1984-spectral,blankenbecler-1985-fractional,andrianov-1986-spectral,stone-1985-elementary,volovik-2003-universe}. In
a gapped system, an adiabatic deformation of the Hamiltonian can only change the
value of this index by closing and reopening the gap. This precisely occurs in
ACs, as we discuss next.

\section{Andreev crystals}
\label{sec:andreev-crystals}

\begin{figure*}[t!]
  \centering \includegraphics[width=.9\linewidth]{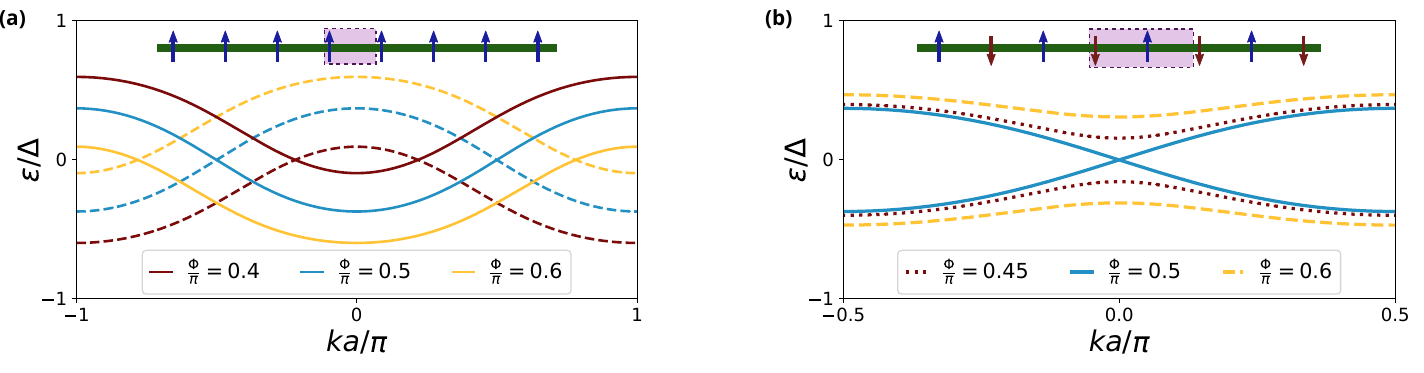}
  \caption{Spectrum of (a) ferromagnetic and (b) antiferromagnetic Andreev crystals for
    energies within the superconducting gap, different values of $\Phi$ and a fixed separation between impurities given by $e^{-a/\xi} = 0.2$. In panel (a) the solid (dashed) line corresponds to spin-up (-down)
    states. The insets  are top views of the structure
    that show the unit cells as shaded regions.}
  \label{fig:spectrum-Andreev-crystals}
\end{figure*}

We define an AC as a periodic arrangement of semiclassical magnetic regions in a
superconductor. In the following we consider a quasi-1D structure of
collinear magnetic regions located at the points $X_n=an$, see
Fig.~\ref{fig:spectrum-Andreev-crystals}, and assume that the lateral dimensions of the system are small enough to treat each conduction channel separately and that the width of the magnetic regions is much smaller than the superconducting coherence length. The latter
allows to treat the magnetic regions in the semiclassical limit as point-like
impurities with a strength proportional to the corresponding magnetic phase
$\Phi_n$, such that the BdG equation for a spin projection $\sigma$ reads
\begin{align}
  \bigg[-i \nu \hbar v_F\hat\tau_3\partial_x + \hat\tau_1\Delta
  - \sigma \hbar v_F \sum_n \Phi_n \delta(x - &X_n)\bigg] \Psi_{\nu\sigma}(x)\nonumber\\
  \label{eq:linearized-BdG-hamiltonian-ugly}
  = \epsilon_{\nu\sigma} &\Psi_{\nu\sigma}(x),
\end{align}
where
$\nu = \pm$ relates to the two electon-hole valleys at $\pm k_F$. 
The key feature of the semiclassical impurities is that quasiparticles do not back-scatter when traversing them, but only accumulate a phase,
\begin{equation}
    \label{eq:BC-semiclassical-imp}
    \Psi_\sigma(X_n^R) = e^{i\sigma\hat\tau_3\Phi_n} \Psi_\sigma(X_n^L).
\end{equation} 
Here, $\sigma$ and $\tau_3$
reflect the fact that the sign of the accumulated phase is different for spin up/down quasiparticles and for electrons/holes, respectively, and $X_n^L$ ($X_n^R$) stands for the position of the left (right) interface of the $n$-th magnetic region. The delta functions in the first order differential equation shown in Eq.~\eqref{eq:linearized-BdG-hamiltonian-ugly} are a shorthand notation of the boundary conditions introduced by the semiclassical impurities, Eq.~\eqref{eq:BC-semiclassical-imp}.
The absence of back scattering allows separate treatment of the two Fermi valleys, so that we can drop the $\nu$ index for brevity. 

The general solution to
Eq.~\eqref{eq:linearized-BdG-hamiltonian-ugly} in the region between two
neighboring impurities, $X_n < x < X_{n+1}$, reads
\begin{equation}
  \label{eq:Bogoliubov-spinor-pm-decomposition}
  \Psi_\sigma(x) = B^{+}_{\sigma n+1} e^{\tfrac{x-X_{n+1}}{\xi}} \ket{+}
  + B^{-}_{\sigma n} e^{-\tfrac{x-X_n}{\xi}} \ket{-}.
\end{equation}
Here $\xi \equiv \frac{\hbar v_F}{\sqrt{\Delta^2 - \epsilon^2}}$ is the
superconducting coherence length, $B_{\sigma n}^{+(-)}$ is the amplitude of the contribution from the spinor that decays from the $n$-th magnetic region to the
left (right), and
\begin{equation}
  \label{eq:basis-vectors}
  \ket{\pm} \equiv \frac{e^{\pm i\theta/2}}{\sqrt{2\cos\theta}} \left(
    \begin{array}{c}
      1 \\ \pm i e^{\mp i\theta}
    \end{array}\right),  
\end{equation}
where $e^{i\theta} \equiv \frac{\sqrt{\Delta^2-\epsilon^2} + i\epsilon}{\Delta}$ is the Andreev factor. Applying the boundary conditions in Eq.~\eqref{eq:BC-semiclassical-imp} to this relations we obtain the equations for the $B^{\pm}$ coefficients in
Eq.~\eqref{eq:Bogoliubov-spinor-pm-decomposition}, which can be recast into an
effective tight binding model by keeping terms up to first order in
$e^{-a/\xi}$. Specifically, in the limit where $e^{-a/\xi}\ll 1$, coefficients $B^-$ at each site $n$ can be related to $B^+$ as follows, 
$$
B_{\sigma n}^- \approx i\sigma \frac{\Delta \sin\Phi_n}{\sqrt{\Delta^2-\epsilon^ 2}} B_{\sigma n}^+,
$$ 
while the rescaled $B^+$ coefficients, $b_{\sigma n} \equiv \sin\Phi_n B_{\sigma n}^+$, satisfy a tight binding type eigenvalue problem,
\begin{equation}
  \label{eq:tight-binding-equations}
  \big(\sigma \omega_\sigma - \omega_{0n}\big) b_{\sigma n} = 
  t_{n+1} b_{\sigma n+1} + t_{n} b_{\sigma n-1}. 
\end{equation}
Here the effective eigenvalue $\omega_{\sigma} \equiv
\frac{\epsilon_{\sigma}}{\sqrt{\Delta^2-\epsilon_{\sigma}^2}}$ is a function of the physical energy $\epsilon_{\sigma}$, $\omega_{0n} =
\frac{\cos\Phi_n}{\sin\Phi_n}$ is the value of $\omega_{\sigma}$ at the energy
$\epsilon_{0n} = \Delta \frac{|\sin\Phi_n|}{\tan\Phi_n}$ of the $n$-th single-impurity
(spin-up) bound state
\cite{andreev-1966-electron,konschelle-2016-semiclassical,konschelle-2016-ballistic},
and $t_{n} \equiv - \frac{e^{-a/\xi}}{\sin\Phi_n}$ is the hopping amplitude.
The expression in Eq.~\eqref{eq:tight-binding-equations} is valid for any AC
with arbitrary distribution of collinear magnetization. Here we focus on two
cases that show rather different qualitative results: the ferromagnetic and
antiferromagnetic ACs described by equal magnetic regions pointing in the same, $\Phi_n = \Phi$, or alternating,
$\Phi_n = (-1)^n\Phi$, directions, respectively.

\subsection{Ferromagnetic ACs with $\Phi_n = \Phi$}
Ferromagnetic ACs are built by equal magnetic regions whose exchange field points in the same direction, and are described by a unit cell containing a single magnetic region.
The solution of Eq.~\eqref{eq:tight-binding-equations} for such systems simply reads $b_{\sigma
  n} = e^{ikna}$ and $\omega_\sigma(k) = \sigma\big(\omega_0 + 2t \cos ka\big)$,
where $k$ is the Bloch momentum \cite{ashcroft-1976-solid}. Hence, the physical
Andreev energy bands are
\begin{equation}
  \label{eq:ferro-dispersion-eq}
  \frac{\epsilon_\sigma(k)}{\Delta} = \sigma
  \frac{\omega_0 + 2t\cos ka}{\sqrt{1 + \big(\omega_0 + 2t\cos ka\big)^2}}\; ,
\end{equation}
where $t$ has to be evaluated at the energy of the single impurity level
$\epsilon_0$.  In Fig.~\ref{fig:spectrum-Andreev-crystals}b we show the
resulting energy spectrum within the Brillouin zone, $-\pi/a<k<\pi/a $, for
different values of $\Phi$. It consists of two symmetric Andreev bands, one for
each spin projection $\sigma$, centered at $\sigma \epsilon_0$. With increase of $\Phi$ the two bands overlap, but remain independent as
they correspond to different spin projections. As long as there is a gap between
the bands, variations of $\Phi$ do not modify the spectrum asymmetry and, thus,
the spin polarization per unit cell remains unchanged [{\it cf.}
Eq. (\ref{eq:spec-asy-spin-polarization-as-spectral-asym})]. When the bands
overlap the spin continuously increases with the further increase of $\Phi$
until the bands pass through each other and the gap reopens.  After reopening
the total spin change is $\hbar/2$ per Fermi valley ($i.e.$, $\hbar$ in
total).

\subsection{Antiferromagnetic ACs with $\Phi_n = (-1)^n\Phi$}

Antiferromagnetic ACs, formed by equal magnetic impurities with the direction of their exchange fields alternatig between up and down along the $z$-axis, show some features of greater interest. In this case the unit cell contains
two anti-aligned impurities (see the inset sketch in
Fig.~\ref{fig:spectrum-Andreev-crystals}c) and it is convenient to rewrite
Eq.~\eqref{eq:tight-binding-equations} as follows:
\begin{equation}
  \label{eq:antiferro-tight-binding}
  \big(\sigma \omega_\sigma - \hat{\Omega}_{0}\big)  C_{\sigma m} =
  \hat T C_{\sigma m-1} + \hat T^\dagger C_{\sigma m+1}\; , 
\end{equation}
where now $C_{\sigma m}\equiv {[b_{\sigma 2m} \ \ b_{\sigma 2m+1}]}^T$ is a spinor, and the
matrices
\begin{equation}
  \label{eq:antiferro-tight-binding-matrices}
  \hat \Omega_{0} = \left(
    \begin{array}{cc}
      \omega_0 & t \\ t & -\omega_0
    \end{array}\right),
  \qquad
  \hat T = \left(
    \begin{array}{cc}
      0 & -t \\ 0 & 0
    \end{array}\right),
\end{equation}
correspond to the unit-cell Hamiltonian and the inter-cell hopping,
respectively. Equation~\eqref{eq:antiferro-tight-binding} describes a chain with
diatomic unit cell and the dispersion relation, $\omega_\sigma =
\pm\sqrt{\omega_0^2 + 4t^2\sin^2 ka}$. This translates into the following Andreev
bands
\begin{equation}
  \label{eq:antiferro-dispersion-eq}
  \frac{\epsilon_\sigma(k)}{\Delta} = \pm \sqrt{
    \frac{\omega_0^2 + 4t^2\sin^2 ka}{1 + \omega_0^2 + 4t^2\sin^2 ka}}\; , 
\end{equation}
shown in Fig.~\ref{fig:spectrum-Andreev-crystals}c.  Because the period is
doubled with respect to the ferromagnetic case, the number of bands is also
doubled.  There are two bands per spin specie which are fully symmetric with
respect to the Fermi energy and, therefore, the spin
polarization is zero [\textit{c.f.} Eq. (\ref{eq:spec-asy-spin-polarization-as-spectral-asym})].  The spectrum shows a gap equal to $2\omega_0=
2\frac{\cos\Phi}{\sin\Phi}$. The gap is finite for all $\Phi$, except for
half-integer values of $\Phi/\pi$, when it closes and the spectrum exhibits a
Dirac point at $k_D=0$. In the vicinity of the critical values,
$\Phi=\pi(l+\frac{1}{2})$, where $l$ is an integer, the eigenvalue problem of
Eq.~\ref{eq:antiferro-tight-binding} linearized around the Dirac point in the
$k$-space reads:
\begin{equation}
  \label{eq:antiferro-Dirac-repr}
  \left(
    \begin{array}{cc}
      \sigma \omega_\sigma - \omega_0 & -2it ka \\
      2it ka & \sigma \omega_\sigma + \omega_0
    \end{array}\right) C_\sigma(k) = 0,
\end{equation}
which has the form of a 1D Dirac equation with $\omega_0$ playing the role of the
mass. The closing and reopening of the gap is associated with a sign change of
the mass term (gap inversion).  Interestingly, the gap can also get inverted
without closing: at values of $\Phi = l\pi$ the Andreev bands merge into the
continuum of the spectrum and reenter in the superconducting gap in inverted
order\cite{rouco-2019-spectral}.

\section{Inverted antiferromagnetic AC junctions}
\label{sec:AC-junctions}

Various realizations of an inhomogeneous Dirac model with the mass inversion
have been widely studied in quantum field theory and condensed matter physics
\cite{jackiw-1976-solitons,su-1980-solitonb,su-1979-solitons,goldstone-1981-fractional,mackenzie-1984-illustrations,stone-1985-elementary,volkov-1985-two}. The
most striking features of this model are the presence of bound states at the
interface where the mass-inversion takes place and the fractionalization of the
interface charge.
As we discuss next, a junction between two antiferromagnetic ACs with inverted
gap is another example of such systems, but with a fractionalized interface
spin, instead of a charge.

\begin{figure}[t!]
  \centering \includegraphics[width=.9\linewidth]{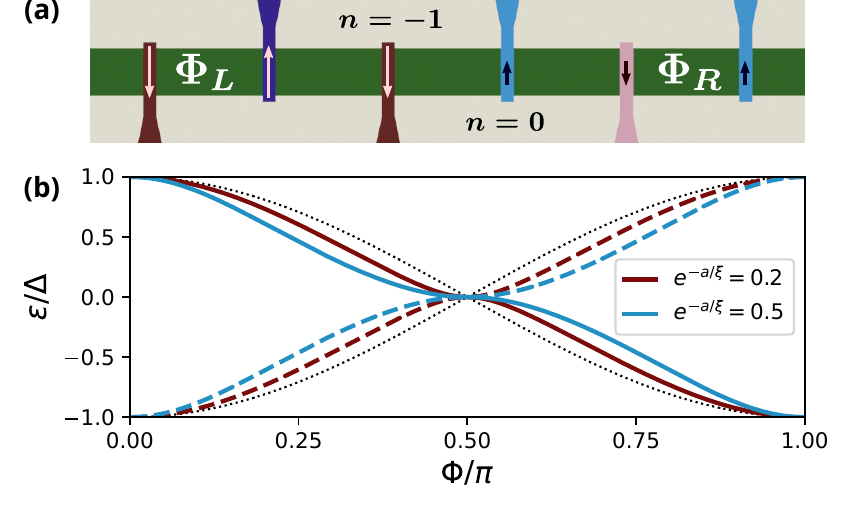}
  \caption{(a) Sketch of a junction between two antiferromagnetic Andreev
    crystals. (b) Energy of the spin up (solid lines) and down (dashed lines)
    bound states in a symmetric inverted junction, $\Phi_R = -\Phi_L = \Phi$, in
    terms of $\Phi$ and for different values of $e^{-a/\xi_0}$
    [Eq.~\eqref{eq:antiferro-juntion-symmetric-energy}]. The dotted black lines
    are the single impurity Andreev levels, $\pm\epsilon_0$, that determine the
    gap edges}
  \label{fig:antiferro-junction}
\end{figure}

To establish the analogy, we consider a junction between two semi-infinite
antiferromagnetic ACs, where the separation between impurities, $a$, remains
constant all along the structure and the magnetic region in the left and right
crystal are described by a magnetic phase equal to $\Phi_L$ and $\Phi_R$,
respectively (see the sketch in Fig.~\ref{fig:antiferro-junction}a). Such a
system is described by the tight-binding equations,
Eq.~\eqref{eq:antiferro-tight-binding}, at each side of the junction, namely
\begin{equation}
  \label{eq:antiferro-junction-tight-binding-left}
  \big(\sigma \omega_\sigma - \hat\Omega_{0L}\big) C_{\sigma m} =
  \hat T_{m-1} C_{\sigma m-1} + \hat T_{m+1}^\dagger C_{\sigma m+1},
\end{equation}
at the left chain ($m < 0$) and
\begin{equation}
  \label{eq:antiferro-junction-tight-binding-right}
  \big(\sigma \omega_\sigma - \hat\Omega_{0R}\big) C_{\sigma m} =
  \hat T_{m-1} C_{\sigma m-1} + \hat T_{m+1}^\dagger C_{\sigma m+1},
\end{equation}
at the right chain ($m \geq 0$). Here, $\hat \Omega_{0L}$ ($\hat \Omega_{0R}$)
stands for the expression of $\hat \Omega_0$ in
Eq.~\eqref{eq:antiferro-tight-binding-matrices} with $\Phi = \Phi_L$ ($\Phi =
\Phi_R$). We look for bound states, {\it i.e.}, solutions that decay as
$C_{\sigma m} =C_\sigma^L e^{m\kappa_\sigma^L}$ into the left crystal and as
$C_{\sigma m} = C_\sigma^R e^{-m\kappa_\sigma^R}$ into the right one, where the
decay is determined by the positive-real-part complex number,
$\kappa_\sigma^{L(R)}$. From
Eqs.~\eqref{eq:antiferro-junction-tight-binding-left} and
\eqref{eq:antiferro-junction-tight-binding-right} we find that
\begin{equation}
  \label{eq:antiferro-junction-L-R-kappa}
  \sinh \frac{\kappa_\sigma^{L(R)}}{2} = \frac{\sqrt{\omega_{0L(R)}^2 - \omega_\sigma^2}}{2|t_{L(R)}|},
\end{equation}
where $t_L = t_{m<0}$ and $t_R = t_{m\geq0}$, and that the bound state exists only
when the following equation
\begin{equation}
  \label{eq:antiferro-junction-secular-equation}
  \frac{\sigma \omega_\sigma - \omega_{0L} }{\sqrt{\omega_{0L}^2 - \omega_\sigma^2}}
  e^{-\tfrac{\kappa_\sigma^L}{2}} = 
  - \frac{\sigma \omega_\sigma - \omega_{0R}}{\sqrt{\omega_{0R}^2 - \omega_\sigma^2}}
  e^{\tfrac{\kappa_\sigma^R}{2}},
\end{equation}
is fulfilled. According to Eq.~\eqref{eq:antiferro-junction-L-R-kappa} a bound
state exists only if $|\omega_{\sigma}| < |\omega_{0L}|$ and $|\omega_{\sigma}| < |\omega_{0R}|$ at the same time.
This implies that Eq.~\eqref{eq:antiferro-junction-secular-equation} has a
solution only in inverted junctions with $\text{sign}(\omega_{0R}) =
-\text{sign}(\omega_{0L})$. The solution is especially simple when $\omega_{0R} =
-\omega_{0L} \equiv \omega_0$ and reads $\omega_\sigma = \sigma
\text{sign}(\omega_0) \left(\sqrt{\omega_0^2 + t^2} - |t| \right)$. This gives
the following physical energy of the bound state,
\begin{equation}
  \label{eq:antiferro-juntion-symmetric-energy}
  \frac{\epsilon_\sigma}{\Delta} = \sigma \text{sign}(\omega_0)
  \frac{\sqrt{\omega_0^2-t^2} - |t|}{\sqrt{1 + \left(\sqrt{\omega_0^2-t^2} - |t|\right)^2}}.
\end{equation}
In Fig.~\ref{fig:antiferro-junction}b we show the bound states for both spin
projections as a function of $\Phi$ for different values of $-t\sin\Phi =
e^{-a/\xi}$. Near the inversion point $|\omega_0/t| \rightarrow 0$,  the
bound states are almost degenerate approaching zero energy, $\epsilon_\sigma
\rightarrow 0$. This is reminiscent of a zero mode in a continuum 1D Dirac model
with mass inversion \cite{jackiw-1976-solitons}. As the bandwidth gets
comparable to $\omega_0$, the states split forming a symmetric pair of levels in the gap between the Andreev bands.

\begin{figure}[t!]
  \centering \includegraphics[width=.9\linewidth]{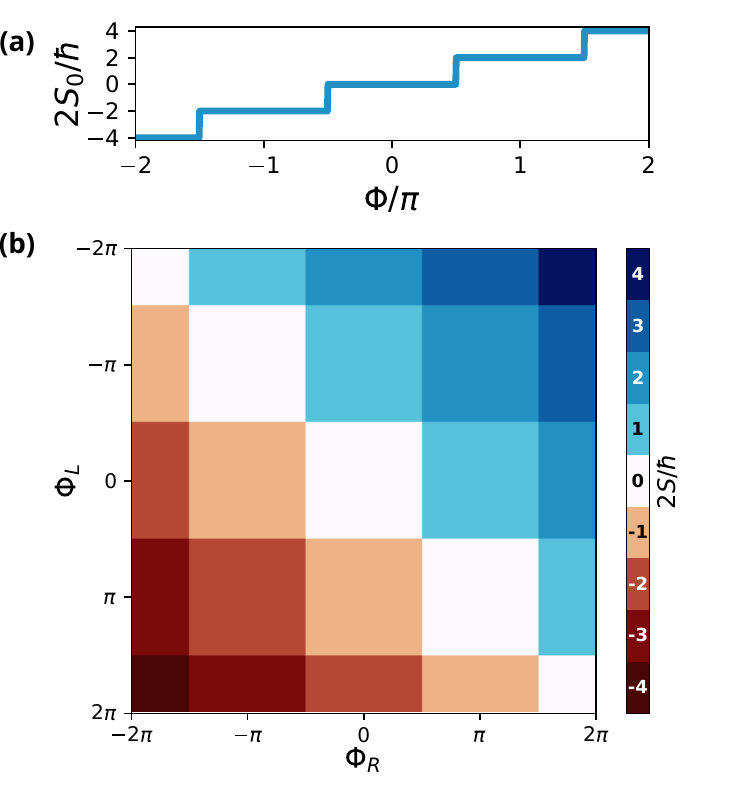}
  \caption{(a) Total spin polarization of the single-impurity system in terms of
    $\Phi$. (b) Spin polarization of the junction between two antiferromagnetic
    Andreev chains as a function of $\Phi_L$ and $\Phi_R$. It is calculated from
    Eq.~\eqref{eq:antiferro-junction-spin-polarization}, with $S_0(\Phi_{L(R)})$
    from panel (a).}
  \label{fig:spin-polarization}
\end{figure}

To calculate the spin, $S$, induced at the contact between the two semi-infinite
antiferromagnetic ACs, we average over all possible terminations of the
chains. This is equivalent to the the so-called \textit{sliding window average}
method (see for example section 4.5 of Ref.~\cite{vanderbilt-2018-berry}), used
to compute the surface charge density by averaging over all possible unit cell
choices.
The calculation is specially simple in the limit when the single-impurity
Andreev states are decoupled from each other, $e^{-a/\xi} \ll 1$. In a previous
work\cite{rouco-2019-spectral}, we show that the spin polarization of a single
semiclassical magnetic impurity of magnetic phase $\Phi$ in a quasi-1D superconducting
wire is $2 S_0(\Phi) / \hbar = 2 \Big[\big(\Phi + \tfrac{\pi}{2}\big) \text{ mod
} \pi\Big]$, where ``mod'' stands for the modulo operation and accounts for a
jump by two electronic spins every time the single-impurity levels cross the
Fermi energy. In Fig.~\ref{fig:spin-polarization}a we show the staircase shape
of $S_0(\Phi)$ in terms of the magnetic phase for a single semiclassical impurity.  We now
consider the junction between the two antiferromagnetic ACs.  It has four
possible ending configurations: whether both chains have the same number of up
and down magnetic impurities, the right (left) Andreev chain has an extra up
(down) magnetic region or both chains are unbalanced. Consequently, the total
spin polarization of the junction, calculated from the average over the four
possible configurations, reads
\begin{equation}
  \label{eq:antiferro-junction-spin-polarization}
  S = \frac{S_0(\Phi_R) - S_0(\Phi_L)}{2}.
\end{equation}
Starting from the uncoupled impurities, if one adiabatically switches on the
coupling, the Andreev bands start widening. However, in the considered
configuration the gap never closes and, as we discussed after
Eq.~\eqref{eq:spec-asy-spin-polarization-as-spectral-asym}, the spin cannot
change and is hence given by
Eq.~\eqref{eq:antiferro-junction-spin-polarization}.  In
Fig.~\ref{fig:spin-polarization}b we show the total spin of the junction in
terms of $\Phi_L$ and $\Phi_R$. Interestingly, the spin polarization can now be
equal to and odd integer times the electronic unit, in contrast to the always
even value of $S_0(\Phi)$. By construction, the half-integer spin (per Fermi
valley) is localized at the junction between ACs. In other words, there is a
fractionalization of the interface spin. Such fractionalization is a local
effect. In a finite system the contribution from the edges will always lead to a
total integer spin per Fermi valley.  Notice that changes on the spin polarization of ACs is
determined by the change of the spectral asymmetry index,
Eq.~\eqref{eq:spec-asy-spin-polarization-as-spectral-asym}, and hence
Eq.~\eqref{eq:antiferro-junction-spin-polarization} is valid beyond the nearest
neighbors tight-binding approximation. This is indeed confirmed by the exact
numerical solution of Eq.~\eqref{eq:linearized-BdG-hamiltonian-ugly}
\footnote{M. Rouco, F. S. Bergeret and I. V. Tokatly, \textit{Article in
    preparation}}
    
\section{Conclussions}

In this work we show that the spin polarization of a gapped system with
collinear magnetization can only change upon gap closing. This occurs in Andreev
crystals for which we present a complete study of their spectral properties for
ferromagnetic and antiferromagnetic configurations.  The spectrum of
antiferromagnetic ACs presents a gap that remains open except for half-integer
values of the magnetic phase $\Phi/\pi$, where a Dirac point is formed. We show
that junctions between antiferrmagnetic ACs with inverted gaps exhibit
interfacial bound states and fractionalization of the surface spin
polarization. We propose realization of these structures using, for example, a
conventional superconducting wire contacted to ferromagnetic fingers such that a
strong periodic exchange field, $h \gg \Delta$, is induced in the
superconductor, see Fig.~\ref{fig:intro-sketches}b. The fingers may
be made of ferromagnetic metals\cite{beckmann-2004-evidence}, like Co or Ni, or
ferromagnetic insulators, like EuS or EuO. The spectrum, and in particular the bound states formed at the interface between two antiferromagnetic chains,   can  be measured by a local tunneling probe, as for example done in Ref. \cite{pillet-2010-andreev}. In case  of a magnetic  probe  one can also  determine the spin-polarization  of such states.

\section*{Acknowledgments}
We acknowledge funding from Spanish Ministerio de Ciencia,
Innovación y Universidades (MICINN) (Projects No. FIS2016-79464-P and
No. FIS2017-82804- P).  I.V.T. acknowledges support by Grupos Consolidados
UPV/EHU del Gobierno Vasco (Grant No. IT1249-19).  The work of F.S.B. is
partially funded by EUs Horizon 2020 research and innovation program under Grant
Agreement No. 800923 (SUPERTED).

\bibliographystyle{apsrev4-1} 
\bibliography{biblist}

\end{document}